\documentclass[a4paper,twocolumn,10pt,final,accepted=2020-12-18]{quantumarticle}
\pdfoutput=1
\usepackage[numbers,sort&compress]{natbib}
\usepackage[utf8]{inputenc}
\usepackage[english]{babel}
\usepackage[T1]{fontenc}
\usepackage{graphicx} \usepackage{amsmath}
\usepackage{amsthm,stmaryrd}
\usepackage{amssymb} \usepackage{physics} 
\usepackage{color}
\usepackage{hyperref}

\usepackage{algorithm}
\usepackage{algpseudocode}

\newcommand{\mb}{\boldsymbol}

\newcommand{\sps}{Stochastic Parameter Shift Rule}

\makeatletter
\newcommand{\StatexIndent}[1][3]{%
  \setlength\@tempdima{\algorithmicindent}%
  \Statex\hskip\dimexpr#1\@tempdima\relax}
\makeatother

\begin{document}

\title{
Measuring Analytic Gradients of General Quantum Evolution with the Stochastic Parameter Shift Rule
} 
 
\author{Leonardo Banchi}
\affiliation{ Department of Physics and Astronomy, University of Florence, via G. Sansone 1, I-50019 Sesto Fiorentino (FI), Italy}
\affiliation{ INFN Sezione di Firenze, via G. Sansone 1, I-50019 Sesto Fiorentino (FI), Italy }
\orcid{0000-0002-6324-8754}

\author{Gavin E.\ Crooks}
\affiliation{X, the moonshot factory (x.company), Mountain View, CA, USA}
\affiliation{Berkeley Institute for Theoretical Science, Berkeley, CA, USA}

\maketitle

\begin{abstract}
	Hybrid quantum-classical optimization algorithms represent one of the most promising 
	application for near-term quantum computers. In these algorithms the goal is to 
	optimize an observable quantity with respect to some classical parameters,
	using feedback from measurements performed on the quantum device. 
	Here we study the problem of estimating the gradient of the function to be optimized 
	directly from quantum measurements,
	generalizing and simplifying some approaches present in the literature, such as 
	the so-called parameter-shift rule. 
	We derive a mathematically exact formula that provides a stochastic algorithm
	for estimating the gradient of any 
	multi-qubit parametric quantum evolution, without the introduction of ancillary qubits 
	or the use of Hamiltonian simulation techniques. The gradient measurement 
	is possible when the underlying device can realize all Pauli rotations in the
	expansion of the Hamiltonian whose coefficients depend on the parameter.
	Our algorithm continues to work, although with some approximations, 
	even when all the available quantum gates are noisy, 
	for instance due to the coupling between the quantum device and an unknown environment. 
\end{abstract}

\section{Introduction}%
\label{sec:introduction}

In the near-term~\cite{preskill2018quantum} quantum computers will be too noisy and the number of operations, 
or {\it depth} of the circuit, will still be too low 
to reliably implement conventional quantum algorithms that require full quantum error correction \cite{montanaro2016quantum}. 
% Near-term quantum computers can create a certain reference state, 
% apply a certain number of parametric quantum gates 
% and then perform a quantum measurement \cite{preskill2018quantum}. 
% %The number of such operations, or {\it depth} of the circuit,
% %is still too low and 
% These operations are still too noisy and their number, 
% or {\it depth} of the circuit, is still too low 
% to reliably implement conventional quantum algorithms 
% \cite{montanaro2016quantum}. 
Therefore, alternative algorithms, 
better suited for exploiting these devices have been proposed, 
such as the variational quantum eigensolver \cite{peruzzo2014variational,jones2019variational},
the quantum approximate optimization algorithm \cite{farhi2014quantum}, 
quantum autoencoders \cite{romero2017quantum}, quantum simulation \cite{li2017efficient},
and quantum classifiers for machine learning 
\cite{schuld2018circuit,benedetti2019parameterized,schuld2018supervised,mitarai2018quantum,farhi2018classification}.
Because of these applications, several companies involved in 
the development of quantum computers 
have released software for the manipulation of parametric quantum 
states \cite{Qiskit,Cirq,bergholm2018pennylane,smith2016practical,broughton2020tensorflow}.

Hybrid quantum-classical optimization algorithms, such 
as the ones mentioned above, try to overcome the 
limitations of current quantum computers by pairing them with a classical device. 
In these hybrid strategies, the ``hard'' part of the algorithm, which typically involves 
the manipulation of objects living in a high-dimensional Hilbert space, is done 
by a quantum computer, which is reset after each measurement.  The classical routine 
then iteratively reprograms the quantum computer in such a way that either the output 
of quantum measurements or the prepared quantum state have the desired property. 
These iterative schemes allow the use of shorter-depth circuits that can be 
implemented within the decoherence time of the device. 
Typically, the manipulation of the quantum state is performed with parametric quantum gates and 
the role of the classical routine is to update those parameters either via 
gradient descent or gradient ascent. Evaluating the gradient of a quantum circuit 
is as hard as the evaluation of the circuit itself, and therefore it is important 
to use the quantum computer for estimating it. Several algorithms have been 
proposed for such purpose, either based on a generalization of the Hadamard test 
\cite{mcclean2018barren,harrow2019low,yuan2019theory,mitarai2019methodology} or on the so-called parameter shift 
rule \cite{mitarai2018quantum,schuld2019evaluating,crooks2019gradients,li2017hybrid}, 
which have a similar complexity. 
Nonetheless, both algorithms can only be applied when the parametric gates 
can be written as $e^{i\theta_t \hat X_t}$, for parameters $\theta_t$, and where
the operators $\hat X_t$ have certain special properties. In the general case one has 
to resort to Hamiltonian simulation techniques \cite{childs2012hamiltonian} that 
increase the complexity of the algorithm.

Here we show that the parameter-shift rule can be generalized to any multi-qubit 
quantum evolution, without the need to introduce any ancillary system or 
Hamiltonian simulation techniques. Our 
generalization is based on a stochastic strategy that is exact in the limit of 
many repetitions of the quantum measurement. We analyse the number of repetitions 
needed to achieve a certain precision by studying the variance 
of our estimation procedure, and numerically observe that it is comparable to that 
of the standard parameter shift rule. 
In near-term computers,
unitary gates are an approximation to a more complex, noisy evolution that couples the 
qubit registers to an unknown environment.
We show that our estimation 
procedure can be applied even when the coupling between system and environment 
cannot be completely suppressed, and when the gates depend on the parameters 
in a complex way. 

Our paper is organized as follows: 
in Sec.~\ref{sec:background_and_notation} we set up the problem and the notation; 
in Sec.~\ref{sec:main_idea} we discuss the main ideas and introduce 
analytical formulae and algorithms for estimating the gradient in the general case; 
in Sec.~\ref{sec:appl} we study 
 applications in quantum control and for optimizing noisy gates;
conclusions are drawn in Sec.~\ref{sec:conc}. 
An alternative deterministic approach based on the Hadamard test is studied in Appendix~\ref{a:hadamard}. 
Explicit pseudo-codes for our algorithms are given in Appendix~\ref{a:explicit}. 
The stochastic variance of our algorithms is studied 
in Appendix~\ref{s:variance}.

\section{Background and notation}%
\label{sec:background_and_notation}

We focus on parametric quantum states  $\ket{\psi(\mb\theta)}$ that depend parametrically on 
$P$ classical real parameters $\{\theta_p\}$ with $p=1,\dots,P$. These states
are obtained by applying a unitary $\hat U(\mb\theta)$ onto a $\mb\theta$-independent 
reference state $\ket{\psi_0}$
\begin{equation}
	\ket{	\psi(\mb\theta)}=\hat U(\mb\theta)\ket{\psi_0}~.
\end{equation}
We study the optimization (either maximization or minimization) of 
the expected value of an observable $\hat C$, taken with respect 
to $\ket{\psi(\mb\theta)}$ 
\begin{equation}
	C(\mb\theta)=\bra{\psi_0}\hat U(\mb\theta)^\dagger \hat C  \hat U(\mb\theta)\ket{\psi_0}~.
	\label{cost}
\end{equation}
Several problems can be mapped to the above optimization, such as variational diagonalization 
and quantum simulation  
\cite{peruzzo2014variational,farhi2014quantum,yuan2019theory}, where $\hat C$ is the Hamiltonian of 
a many-body system and the task is to variationally approximate its ground state;
and quantum state synthesis, where $\hat C=|{\psi_{\rm target}}\rangle\langle{\psi_{\rm target}}|$, 
or some machine-learning classifiers 
\cite{schuld2018circuit,mcclean2018barren}.
Even quantum control problems \cite{khaneja2005optimal} or the simulation of gates with 
time-independent Hamiltonians \cite{banchi2016quantum,innocenti2020supervised} can be written in 
the form~\eqref{cost}. Indeed, consider 
the task of finding a good approximation of a certain target unitary gate $\hat G$ with a 
parametric unitary $\hat U(\mb\theta)$. We may 
define $\ket{\psi(\mb\theta)} = \hat\openone\otimes \hat
U(\mb\theta)\ket{\Phi}$, where $\ket{\Phi}=\sum_{i=1}^d\ket{i,i}/\sqrt{d}$ and
$d$ is the dimension of the Hilbert space, and similarly $|{\psi_{\rm
target}}\rangle = \hat \openone\otimes \hat G\ket{\Phi}$. Then, from \eqref{cost} with 
$\hat C=|{\psi_{\rm target}}\rangle\langle{\psi_{\rm target}}|$, we find 
\begin{equation}
	C(\mb\theta) = \left(\frac{|\Tr \hat G^\dagger \hat U(\mb\theta)|}d\right)^2~,
\end{equation}
which is the function normally maximized in quantum control
problems~\cite{khaneja2005optimal,banchi2016quantum}.

Any unitary operator can be expressed as a matrix exponential 
$
	\hat U(\mb\theta) = e^{i \hat X(\mb\theta)},
$
where $\hat X(\mb\theta)$ is a Hermitian operator. When the unitary 
$	\hat U(\mb\theta) $ is a composition of $T$ simpler gates 
$\hat{U}_t(\mb\theta) $, then we write 
\begin{align}
	\hat U(\mb\theta) &= \prod_{t=1}^T U_t(\mb\theta),
										&	
	\hat U_t(\mb\theta) &= e^{i \hat X_t(\mb\theta)}~,
	\label{Uprod}
\end{align}
where the products are ordered as $\prod_{t=1}^T \hat U_t :=\hat U_T\cdots \hat U_1$.
The 
products of Pauli matrices $\hat \sigma_{\mb \nu} = \hat{\sigma}_{\nu_1}\otimes \cdots 
\otimes \hat{\sigma}_{\nu_N}$ form a basis for the space of $N$-qubit Hermitian operators
\cite{nielsen2006quantum}, where $\mb\nu=(\nu_1,\dots,\nu_N)$ is a multi index, 
$\nu_j$ is either $\{0,x,y,z\}$ and $\hat \sigma_0:=\hat\openone$,
$\hat\sigma_x$, $\hat\sigma_y$, $\hat\sigma_z$ are the Pauli matrices. 
As such, we may expand the operators $\hat X_t(\mb\theta)$ onto this basis and write 
\begin{equation}
	\hat X_t(\mb\theta) = \sum_{\mb\nu} x_{t,\mb\nu}(\mb\theta) \, \hat\sigma_{\mb\nu}~,
	\label{expansion}
\end{equation}
%	\vspace{4mm}
with coefficients $x_{t,\mb\nu}({\mb\theta})=\Tr[\hat X_t(\mb\theta)\hat\sigma_{\mb\nu}]/2^N$.
It is common to restrict attention to gates that only have a single element in the expansion 
\eqref{expansion}, i.e.~$x_{t,\mb\mu}(\mb\theta)=\theta_{t}\delta_{\mb\mu,\mb\nu(t)}$, 
where $\mb\nu(t)$ specifies the kind of parametric gate applied at time $t$, or,
more generally, to gates  with $x_{t,\mb\mu}(\mb\theta)=\theta_t x_{t,\mb\mu}$, for which
\begin{align}
%\hat U_t^{\rm simple} = \exp(i\theta_{t}\,\hat \sigma_{\mb\nu(t)})~,
	\hat U_t^{\rm simple} &= e^{i\theta_t \hat H_t}~,
												& \hat H_t = \sum_{\mb\nu} x_{t,\mb\nu}\hat \sigma_{\mb\nu}~,
	\label{Usimple}
\end{align}
where the operator $\hat H_t$ is independent on the parameters $\mb\theta$. 
Moreover, 
most often we consider gates that act on either 
one- or two-qubit, so at most two Pauli matrices in the product 
$\hat{\sigma}_{\nu_1}\otimes \cdots \otimes \hat{\sigma}_{\nu_N}$ are different from 
the identity. % $\hat\openone$. 
Gates as in Eq.~\eqref{Usimple} are quite common, as they physically correspond to solutions of a
Schr\"odinger equation with Hamiltonian $\hat H_t$ and time parameter $\theta_t$. Yet, they do not 
model the most general physical evolution, e.g.~where the parameters are different from a ``time'', 
which is discussed in this paper. 

When the parametrization is such that all gates can be expressed as a sequence \eqref{Uprod} 
of elementary gates as in in Eq.~\eqref{Usimple}, then the derivative of each gate with respect to its 
parameter is straightforward, as 
$\partial \hat U_t/\partial \theta_t = i \hat H_t \hat U_t$. By exploiting the above identity 
into \eqref{cost}, different approaches have been proposed to evaluate the gradient of $C(\mb\theta)$ 
via a carefully designed quantum circuit and classical post-processing, for
instance using the Hadamard test \cite{harrow2019low,yuan2019theory} or the
parameter shift rule \cite{mitarai2018quantum,schuld2019evaluating}.
The parameter-shift rule can only be applied when $\hat H_t$ has two distinct eigenvalues, 
for instance when there is only one non-zero element in the Pauli expansion \eqref{Usimple},
whereas the Hadamard test is more general but requires controlled operations and ancillary qubits.
Nonetheless, recently a generalization of the Hadamard test without the use of controlled operations 
has also been proposed \cite{mitarai2019methodology}. Moreover, in
Ref.~\cite{crooks2019gradients} the parameter shift rule was also generalized 
to some particular cases where there are more than one term in the expansion \eqref{Usimple}.
However, finding the gradient in the general case for possibly many-body gates was still an open question. 
In the next section we show that by mixing standard operator derivative techniques 
\cite{wilcox1967exponential} with Monte Carlo strategies, we can define a procedure to 
measure gradients of any $C(\mb\theta)$, as in Eq.~\eqref{cost}, with near-term quantum 
hardware. 
While we are unaware of any published work showing that the Hadamard test can
be used beyond the class of simple circuits \eqref{Usimple}, in Appendix~\ref{a:hadamard} we remark that 
an alternative deterministic scheme involving the Hadamard test is possible whenever the parametric gate 
acts on few qubits.

In this paper we study a method to find gradients of general quantum evolution, not restricted 
to gates as in Eq.~\eqref{Usimple}. For instance, our methods allow the computation of gradients 
with gates $e^{i(\hat X_t + \theta_t \hat V_t)}$, even when $[\hat H_t,\hat V_t]\neq 0$, 
for which neither the standard parameter shift rule,
nor the Hadamard test can be applied. Such evolutions may arise for instance when some terms in the 
Hamiltonian may not be completely switched off, e.g.~in a noisy setting.

%On the other hand, in this paper we do not restrict 
%ourselves to the case \eqref{Usimple} and 

We consider the general 
parametrization \eqref{Uprod} with the expansion \eqref{expansion}. 
Via the Leibniz rule, we may write the derivative of the expectation value \eqref{cost} as 
\begin{equation}
	\frac{\partial C(\mb\theta)}{\partial \theta_p}  = \sum_{t,\mb\nu}
	\frac{\partial C}{x_{t,\mb\nu}} \;\frac{\partial
		x_{t,\mb\nu}(\mb\theta)}{\partial \theta_p}~.
		\label{leibniz}
\end{equation}
Thanks to the Leibniz rule \eqref{leibniz},
we may fix $t$ and $\mb\nu$, and %focus on a specific value of $t$ and and 
study the 
derivative of $C$ with respect to $x_{t,\mb\nu}$. % for all possible values of $\mb\nu$.
By repeating the analysis for each 
possible values of $t$ and $\mb\nu$, from Eq.~\eqref{leibniz} 
we may obtain the derivatives with respect to the parameters $\theta_p$,
and hence the gradient. %  via Eq.~\eqref{leibniz}. 
%With a similar argument, we may also fix $\mb\nu$.
Therefore, we fix $t$ and $\mb\nu$ and,
to simplify the notation, we 
drop the dependence on $t$ and $ \mb\nu$ to write
\begin{align}
	x &:= x_{t,\mb\nu}, & 
	\hat V &:= \sigma_{\mb\nu},
				 &
	\hat H &:= \sum_{\mb\mu\neq\mb\nu} x_{t,\mb\mu} \hat \sigma_{\mb\mu}~.
	\label{xdef}
\end{align}
With a similar spirit, we also define 
\begin{align}
	\ket{\phi} &:= \prod_{s=1}^{t-1}\hat U_s\ket{\psi_0}~,	
							 &
							 \hat A := \hat U_{t+}^\dagger \hat C \hat U_{t+}~,
							 \label{prep}
\end{align}
where $\hat U_{t+} = \prod_{s=t+1}^T \hat U_{s}$. Thanks to the above simplified notation,
we may write the function $C$ in \eqref{cost} 
as a function of $x\equiv x_{t,\mb\nu}$ for  {\it fixed} $t$ and $\mb\nu$
\begin{equation}
	C(x) = \bra\phi e^{-i (\hat H+x \hat V)} \,\hat A\, e^{i (\hat H+x \hat V)} \ket\phi~,
	\label{cost_simple}
\end{equation}
all the other terms in \eqref{cost_simple} do not explicitly depend on $x\equiv x_{t,\mb\nu}$.
%\begin{equation}
%	C(\mb x_t) = \bra{\psi_t} \hat U_t^\dagger(\mb x_t) \hat C_t\hat U_t(\mb x_t)\ket{\psi_t}~,
%	\label{costX}
%\end{equation}
%where 
%$\mb x_t = \{ x_{t,\mb\nu} {\rm~for~all~}\nu \}$ 
%\begin{align}
%	\ket{\psi_t} &= \prod_{s=1}^{t-1}\hat U_s\ket{\psi_0}~,	
%							 &
%							 \hat C_t = \hat U_{t+}^\dagger \hat C \hat U_{t+}~,
%\end{align}
%where $\hat U_{t+} = \prod_{s=t+1}^T \hat U_{s}$.  Therefore, 
In other words, 
Eq.~\eqref{cost_simple} is equivalent to 
Eq.~\eqref{cost}, where we have separated the terms that depend on 
$x\equiv x_{t,\nu}$ for fixed $t$ and $\mb\nu$ from the others.

\section{Stochastic Parameter Shift Rule}%
\label{sec:main_idea}
Without loss of generality, we fix $t$ and $\mb\nu$ as described in the previous section, and study 
the derivative of $C(x)$ defined in \eqref{cost_simple}. The derivative with respect to the 
parameters $\theta_p$ can be obtained from \eqref{leibniz} by repeating the analysis for 
all $t$ and $\mb\nu$.  We remark that in Eq.~\eqref{cost_simple} 
the state $\ket{\phi}$ and the operators $\hat H$, $\hat A$, $\hat V$ explicitly depend on $t$, $\mb\nu$, and 
on the other values $x_{t',\mb\nu'}$ with either $t'\neq t$ or $\mb\nu'\neq \mb\nu$, but we 
omit this dependence to simplify the notation. Full algorithms are shown in 
Appendix~\ref{a:explicit}. 

\begin{figure}[t]
\begin{algorithm}[H]
	\caption{Parameter Shift Rule}
  \label{alg:ps}
   \begin{algorithmic}[1]
		 \State initialize the computer in the state $\ket{\phi}$, following the 
		 preparation routine \eqref{prep};
		 \State apply  the gate $e^{i \left(x+\frac\pi{4u}\right) \hat V}$;
		 \State measure the observable $\hat A$ from \eqref{prep} and call the result $r_+$.
		 \State Repeat steps 1 to 3, but on point 2 apply 
		 $e^{i \left(x-\frac\pi{4u}\right) \hat V}$ rather than $e^{i \left(x+\frac\pi{4u}\right) \hat V}$;
		 \State measure $\hat A$ and call the result $r_-$. 
		 \State the sample $g_{t,\mb\nu}=u(r_+-r_-)$ is such that $\partial C/\partial x_{t,\mb\nu} = \mathbb{E}[g_{t,\mb\nu}]$.
   \end{algorithmic}
\end{algorithm}
	\centering
	\includegraphics[scale=0.9]{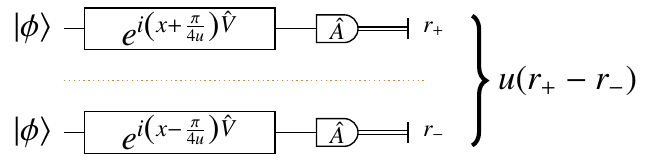}
	\caption{Parameter Shift Rule \cite{mitarai2018quantum,sweke2019stochastic,schuld2019evaluating}, 
		 only applicable to parametric gates as in Eq.~\eqref{Usimple} or, more generally, to
		 parametrizations $e^{i x \hat V}$ where $\hat V$ has two distinct eigenvalues $\pm u$.
		 When $\hat V$ is a product of Pauli matrices as in \eqref{Usimple}, $u=1$. 
		  In the algorithm we consider the derivative
		 $\partial_x C(x)$ of Eq.~\eqref{cost_simple}, when $\hat H=0$. 
	}
	\label{fig:ps}
\end{figure}

\begin{figure}[t]
\begin{algorithm}[H]
  \caption{\sps }
  \label{alg:sps}
   \begin{algorithmic}[1]
		 \State Sample $s$ from the uniform distribution in [0,1];
		 \State initialize the computer in the state $\ket{\phi}$, following the 
		 preparation routine \eqref{prep};
		 \State apply  the gate $e^{i (1-s)(\hat H+x \hat V)}$, namely where parameters $x_{t,\mb\mu}$
		 for fixed $t$ and all possible values of $\mb\mu$ have been rescaled by a factor $(1-s)$; 
		 \State apply the gate $e^{i\pi \hat V/4}\equiv e^{i\pi \hat \sigma_{\mb\nu}/4}$;
		 \State apply  the gate $e^{i s(\hat H+x \hat V)}$, where parameters $x_{t,\mb\nu}$
		 for fixed $t$ and all possible values of $\mb\nu$ have been rescaled by a factor $s$; 
		 \State measure the observable $\hat A$ from \eqref{prep} and call the result $r_+$.
		 \State Repeat steps 2 to 5, but on point 4 apply $e^{i\pi \hat
		 \sigma_{\mb\nu}/4}$ rather than $e^{-i\pi \hat \sigma_{\mb\nu}/4}$;
		 \State measure $\hat A$ and call the result $r_-$. 
%		 \For{$q=\{+,-\}$}
%		 \State initialize the computer in the state $\ket{\phi}\equiv \ket{\psi_t}$, following the 
%		 preparation routine \eqref{prep};
%		 \State apply  the gate $e^{i (1-s)(\hat H+x \hat V)}$, namely where parameters $x_{t,\mb\mu}$
%		 for fixed $t$ and all possible values of $\mb\mu$ have been rescaled by a factor $(1-s)$; 
%		 \State apply the gate $e^{iq\pi \hat V/4}\equiv e^{\pm i\pi \hat \sigma_{\mb\nu}/4}$https://www.overleaf.com/project/5e7b92ba24c7790001dc8d22
%		 where the sign depends on $q$;
%		 \State apply  the gate $e^{i s(\hat H+x \hat V)}$, where parameters $x_{t,\mb\nu}$
%		 for fixed $t$ and all possible values of $\mb\nu$ have been rescaled by a factor $s$; 
%		 \State measure the observable $\hat A$ from \eqref{prep} and call the result $r_q$.
%		 \EndFor
		 %\State An estimate of $\partial C/\partial x_{t,\mb\nu}$ is given by $r_+-r_-$. 
		 \State the sample $g_{t,\mb\nu}=r_+-r_-$ is such that 
		 $\partial C/\partial x_{t,\mb\nu} =\mathbb{E}[g_{t,\mb\nu}]$. 
   \end{algorithmic}
\end{algorithm}
	\centering
	\includegraphics[scale=0.85]{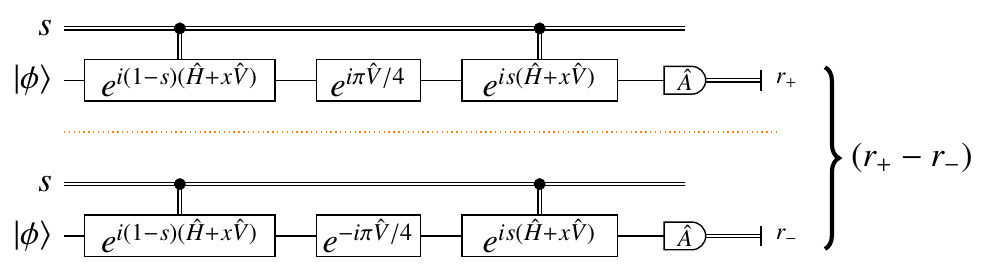}
	\caption{\sps, valid for any operator $\hat H$. In the picture, 
	the values of the classical parameter $s$ are the same.} 
	\label{fig:sps}
\end{figure}

The main tool behind our analysis is the following operator identity \cite{wilcox1967exponential} 
\begin{equation}
	\frac{\partial e^{\mathcal Z}}{\partial x} = \int_0^1 ds\, e^{s {\mathcal Z}} 
	\frac{\partial {\mathcal Z}}{\partial x} e^{(1-s) {\mathcal Z}}~,
	\label{id}
\end{equation}
which is valid for any bounded operator $\mathcal Z$. We may rewrite
Eq.~\eqref{cost_simple} as 
\begin{equation}
	C(x) = \Tr(\hat A e^{\mathcal Z}[\hat{\rho}])~,
	\label{costZ}
\end{equation}
for $\hat \rho=\ket\phi\!\bra\phi$ and for a superoperator 
%\begin{align}
	$
	\mathcal Z[\hat \rho] := [ i(\hat H + x\hat V) ,\hat \rho],
	$
%\end{align}
where $[\hat A,\hat B]=\hat A\hat B-\hat B\hat A$. 
Eq.~\eqref{costZ} then follows from Baker–Campbell–Hausdorff identity
$e^{[\hat X,\cdot]}\hat Y=e^{\hat X}\hat Y e^{-\hat X}$
\cite{miller1973symmetry}.
We also introduce the superoperator 
\begin{equation}
	\mathcal V:=
	\frac{\partial \mathcal Z}{\partial x} = i [\hat V,\cdot].
	\label{Zderiv}
\end{equation}
Now we focus on the exponential 
$e^{\lambda \mathcal V} $ with $\mathcal V$ defined in \eqref{Zderiv}. 
From series expansion, since $\hat V$ is a tensor product of Pauli matrices \eqref{xdef} and,
as such, $\hat V^2=\hat\openone$, it is simple to show that 
\begin{equation}
	e^{\lambda \mathcal V}[\hat \rho]  =\hat  \rho + \sin^2(\lambda)(\hat V\hat
	\rho \hat V-\hat \rho) + \frac{i}2 \sin(2\lambda) [\hat V,\hat \rho]~,
\end{equation}
from which we get by explicit computation
\begin{equation}
	\mathcal V
	e^{\lambda \mathcal V}[\hat \rho]  \equiv
	\frac{\partial e^{\lambda \mathcal V}}{\partial \lambda}[\hat \rho] = 
	e^{(\lambda+\pi/4)\mathcal V}[\hat \rho] - 
	e^{(\lambda-\pi/4)\mathcal V}[\hat \rho]~.
	\label{Vderiv} \end{equation}
When $\hat H\equiv 0$, it is  $\mathcal Z=x \mathcal V$ and we may use the
above equation with $\lambda=x$ to take derivatives in \eqref{costZ}. As 
a result, we get $\partial_x C(x) = C(x+\pi/4)-C(x-\pi/4)$, which is the so-called 
parameter shift rule, described in Fig.~\ref{fig:ps},
often used for training quantum circuits 
\cite{mitarai2018quantum,sweke2019stochastic,schuld2019evaluating}.
Note that,
with the formalism of the previous section, $\hat H=0$ corresponds to the use of 
the simpler parametric unitaries of Eq.~\eqref{Usimple}. A more general version of 
the parameter shift rule can be obtained when the operator~$\hat V$ has only
two distinct eigenvalues
\cite{mitarai2018quantum,schuld2019evaluating}.  Indeed, we note that the only property we used in \eqref{Vderiv} 
is $\hat V^2=\hat\openone$, which is true for any product of Pauli matrices. 
If $\hat V$ has only two possible eigenvalues $c\pm u$, then we may write $\hat V=u \hat V' + c 
\hat \openone$ where $\hat V'^2=\hat\openone$ and the dependence on $c$ disappears in \eqref{Zderiv}. 
Therefore, it is straightforward to generalize the above derivation and find 
$
\partial_x C(x) = u \left[C\left(t+\frac{\pi}{4u}\right)-  C\left(t-\frac{\pi}{4u}\right)\right]
$.
The resulting algorithm is described in Fig.~\ref{fig:ps}.
Although the parameter shift rule can be made slightly more general, 
for instance by replacing the operator $\hat\sigma_{\mb\nu(t)}$ in \eqref{Usimple} with 
another operator that has, like  $\hat\sigma_{\mb\nu(t)}$, only two 
possibly degenerate eigenvalues, 
it cannot be applied in the general case where 
%Note that that the algorithm from Fig.~\ref{fig:ps} cannot be applied when 
$\hat H \neq 0$. Nonetheless, we show that the %algorithm from Fig.~\ref{fig:ps} 
parameter shift rule 
can be generalized by combining Eq.~\eqref{Vderiv} with Eq.~\eqref{id}.  Indeed, evaluating 
\eqref{Vderiv} for $\lambda=0$ we get 
\begin{equation}
	\frac{\partial \mathcal Z[\hat \rho]}{\partial x} = 
	i[\hat V,\hat \rho] = e^{i \pi \hat V/4}\hat \rho e^{-i \hat V\pi/4} - e^{-i \pi \hat V/4}\hat \rho e^{i \hat V\pi/4}~.
	\label{commV}
\end{equation}
From the above equation, calling 
\begin{align}
	C_\pm(x,s) &= \bra\phi U_\pm(x,s)^\dagger \,\hat A\, U_\pm(x,s) \ket\phi~,
	\label{costp}
	\\
	U_\pm(x,s) &=  e^{i s(\hat H+x \hat V)}  e^{\pm i \frac \pi4\hat V} 
	e^{i (1-s)(\hat H+x \hat V)}~,
	\label{Upm}
\end{align}
we get from \eqref{id} and \eqref{commV} 
\begin{equation}
	\partial_x C(x) = \int_{0}^1  [ C_+(x,s) - C_-(x,s)]\,ds~.
	\label{cost_deriv}
\end{equation}
Eqs.\eqref{costp}-\eqref{cost_deriv} represent the central result of this paper. 
Thanks to those formulae, we introduce the \sps, shown in 
Fig~\ref{fig:sps}.
We now give some more details on how the \sps~directly follows from Eq.~\eqref{cost_deriv}. 
Indeed, let $\hat A = \sum_m a_m \ket m\!\bra m$ be the eigenvalue 
decomposition of $\hat A$. Then, because of  the Born rule,
the outcomes $r_\pm$ are one of the possible values $a_m$ 
with probability 
\begin{equation}
	p_\pm(m|s) = |\bra m U_\pm(x,s)\ket\phi|^2~.
	\label{probms}
\end{equation}
Taking the expectation value with respect to the measurement outcomes and with respect to the uniform 
probability over $s$, since 
$\partial_x C(x) = \sum_m a_m \int_s [p_+(m|s)-p_-(m|s)]ds$,
we get from Eq.~\eqref{cost_deriv} 
\begin{equation}
	\mathbb{E}[r_+-r_-] = \partial_x C(x)~.
	\label{unbiased}
\end{equation}
For a single measurement, both the Parameter Shift
Rule of Fig.~\ref{fig:ps} and the \sps~of
Fig.~\ref{fig:sps} provide a random difference between two eigenvalues of $\hat A$. 
Only in the limit over many repetitions of those algorithms does the average over 
the outcomes converge to the exact value of the gradient. 

We now study how many repetitions are 
needed to estimate the gradient with a given accuracy. 
Due to the Chebyshev inequality, 
the number of repetitions to achieve a certain precision depends on the variance
of the random outcomes. Since the outcomes $r_\pm$ are independent, the variance of the 
estimator is ${\rm Var}(r_+-r_-) = {\rm Var}(r_+) + {\rm Var}(r_-)$ with 
${\rm Var}(r_\pm) = \int_0^1 \sum_m a_m^2 p_\pm(m|s) ds - \mathbb{E}[r_\pm]^2 
\leq \max_m a_m^2 = \|\hat C\|^2_\infty$. In particular, if $\hat C = \sum_{\mb\nu} c_{\mb\nu} 
\hat\sigma_{\mb\nu}$, then we can bound the variance as 
${\rm Var}(r_+-r_-) \leq 2\sum_{\mb\nu} c_{\mb\nu}^2$. The same bound can be obtained for 
the standard parameter shift rule so, in spite of the further sampling  over $s$, 
the stochastic parameter shift rule has the same worst case performance of the standard 
parameter shift rule. Moreover, according to \cite{van2020measurement}, 
the variance obtainable with the Hadamard test is bounded by $n_t \sum_{\mb\nu} c_{\mb\nu}^2$, 
where $n_t$ is the number of non-zero Pauli operators in the expansion \eqref{Usimple},
so it is slightly lower for $n_t=1$.

For any generic parametrization, 
the variance of the derivative can then be estimated from 
the Leibniz rule \eqref{leibniz}: for instance, if $c_{\mb\nu} = \mathcal O(1)$ and 
$\partial x_{t,\mb\nu}/\partial \theta_p = \mathcal O(1)$, then assuming independent 
measurements, the variance of the estimator can be bounded as $\mathcal O(2 n_p n_c)$ where 
$n_p$ is the number of non-zero $\partial x_{t,\mb\nu}/\partial \theta_p$ and $n_c$ is the 
number of non-zero $c_{\mb\nu}$. 
The above considerations apply for upper bounds on the variance.
On the other hand, in Appendix~\ref{s:variance} we numerically study the variance 
of the gradient estimator obtained with the \sps~and show that it is comparable with 
that of the standard Parameter Shift Rule.

%compare the variance of the gradient 
%estimators obtained via either Parameter Shift Rule or our generalization of 
%Fig.~\ref{fig:sps}. 

\subsection{Stochastic optimization}\label{sec:stoc_opt}

In the previous section we have introduced an algorithm (Fig.~\ref{alg:sps}) 
to use a quantum computer to sample from a random variable whose average 
is equal to the gradient of a certain circuit. We say that the output 
of the \sps~provides an {\it unbiased estimator} of the gradient, in the sense of Eq.~\eqref{unbiased}. 

We now focus on the original problem, namely a parametric unitary \eqref{Uprod} with 
many parameters as in \eqref{expansion}. We can use the algorithm of Fig.~\ref{fig:sps} 
to sample $g_{t,\mb\nu}$ with the property $\partial C/\partial
x_{t,\mb\nu}=\mathbb{E}[g_{t,\mb\nu}]$. 
By repeating the procedure many times and with all possible values of $t$ and $\mb\nu$,
due the linearity of the Leibniz rule~\eqref{leibniz}, we may write 
\begin{equation}
	\frac{\partial C(\mb\theta)}{\partial \theta_p}  = \mathbb{E}\left[\sum_{t,\mb\nu}
	{g_{t,\mb\nu}} \;\frac{\partial
	x_{t,\mb\nu}(\mb\theta)}{\partial \theta_p}\right]~,
		\label{stocgrad}
\end{equation}
where the expectation value $\mathbb{E}$ has the same meaning as in Eq.~\eqref{unbiased}. 
The full algorithm is shown in Appendix~\ref{a:explicit}, algorithm~\ref{alg:stocgrad}.
The problem with this approach is that we have to repeat algorithm~\ref{alg:sps} many 
times, each time resetting the quantum machine, to get a single sample. 

We now introduce a simpler 
unbiased estimator of the gradient that requires significantly
fewer operations to get a single sample. A similar technique has been developed in 
\cite{harrow2019low,sweke2019stochastic} 
for parametrizations as in Eq.~\eqref{Usimple}, which was dubbed 
{\it doubly stochastic} gradient descend. Here we generalize that approach to general 
quantum evolution, as in Eq.~\eqref{Uprod}. 
We start by defining a probability distribution from the ``weights'' 
$\partial_p x_{t,\mb\nu}(\mb\theta)$, where $\partial_p\equiv
\frac{\partial }{\partial \theta_p}$,  as 
%$\frac{\partial x_{t,\mb\nu}(\mb\theta)}{\partial \theta_p}$ as 
\begin{align}
	q_p({t,\mb\nu}) &=\frac1{\mathcal N} 
	\left|\frac{\partial x_{t,\mb\nu}(\mb\theta)}{\partial \theta_p}\right| ~,
							 &
							 \sum_{t,\mb\nu}q_p({t,\mb\nu}) = 1~,
							 \label{probtnu}
\end{align}
with $\mathcal N = \sum_{t,\mb\nu} |\partial_p x_{t,\mb\nu}(\mb\theta)|$. Setting 
$n_{p,t,\mb\nu} = \mathcal N {\rm sign}\left(\partial_p x_{t,\mb\nu}(\mb\theta)\right)$
%\sqrt{\sum_{t,\mb\nu}\left(\frac{\partial x_{t,\mb\nu}(\mb\theta)}{\partial \theta_p}\right)^2}$,  
	we may then write Eq.~\eqref{stocgrad} as 
\begin{equation}
	\frac{\partial C(\mb\theta)}{\partial \theta_p}  = \mathbb{E}_{(t,\mb\nu)\sim q_p}\left[
n_{p,t,\mb\nu}\;
	\mathbb{E}\left(
	{g_{t,\mb\nu}} \right)\right]~,
		\label{dstocgrad}
\end{equation}
where $\mathbb{E}_{(t,\mb\nu)\sim q}$ means that, at each iteration, 
$t$, and $\mb\nu$ are sampled from the distribution \eqref{probtnu}. 
When the functional dependence on the parameters is known, 
all quantities $q_p(t,\mb\nu)$ and $n_{p,t,\mb\nu}$ 
can be easily computed at each iteration without having to deal with 
exponentially large spaces. 
The above equation \eqref{dstocgrad} 
allows us to define a simple ``doubly stochastic'' 
gradient estimator via the following rule
   \begin{algorithmic}[1]
		 \State sample $t$ and $\mb\nu$  from the distribution \eqref{probtnu};
		 \State use Algorithm~\ref{alg:sps} to get an estimate $g_{t,\mb\nu}$;
		 \State the sample $r_{p,t,\mb\nu}=g_{t,\mb\nu} n_{p,t,\mb\nu}$ is such that 
		 $\partial C/\partial \theta_p = \mathbb{E}[r_{p,t,\mb\nu}]$.
   \end{algorithmic}
%	 where the last line follows from Eq.~\ref{dstocgrad}. 
The full algorithm is shown in Appendix~\ref{a:explicit}, algorithm~\ref{alg:dstocgrad}. Based on the above equation,
in Appendix~\ref{a:explicit}
we also define an algorithm that can provide an unbiased sample 
with a single initialization of the quantum device, algorithm~\ref{alg:ddstocgrad}. 

To summarize the results of this section, we can use either \eqref{stocgrad} or \eqref{dstocgrad} to 
estimate the gradient of an expectation value~\eqref{cost} with a quantum computer. 
Once we have an estimate of the gradient, we can optimize $C(\mb\theta)$ using 
stochastic gradient descent (or ascent) algorithms \cite{bubeck2015convex},
such as Adam \cite{kingma2014adam}. These algorithms are classical, in the sense 
that, given certain parameters $\mb\theta$ and an estimate of the gradient $\mb g$, 
the parameters are updated as $\mb\theta\to\mb\theta\pm\eta \mb g$ for a suitably 
small {\it learning rate}~$\eta$. 
Therefore, we can use a {\it hybrid} quantum-classical approach to optimize $C(\mb\theta)$ 
where the hard calculations, namely the estimation of the gradients, are delegated to a 
quantum computer, while the update of the parameters is performed classically.

\subsection{Quantum gates with unavoidable drift} 

Depending on the hardware, the application of the gates 
$e^{\pm i\pi \hat V/4}$ in Algorithm~\ref{alg:sps} might be problematic. 
Let us consider a quantum computer that can only apply the 
parametric gates 
\begin{equation}
	\hat U(t,b) = e^{i t (\hat H_0 + b \hat H_1)}~,
	\label{withdrift}
\end{equation}
where $\hat H_0$ is some {\it drift} Hamiltonian that cannot be completely switched off, aside from 
the trivial case $t=0$. Such ``simple'' device is still capable of 
universal quantum computation, provided that the operators $\hat H_0$ and $\hat
H_1$ are multi-qubit operators that generate the full Lie algebra \cite{lloyd1996universal}. 
Here though, for simplicity, we consider the case where both $\hat H_0$ and $\hat H_1$ are 
tensor products of Pauli operators, as introduced in Sec.~\ref{sec:background_and_notation}. 
The parameters in the above gate are $\mb\theta=(t,b)$. Using  the notation 
of Eq.~\eqref{expansion} we may write
\begin{equation}
	\hat U(t,b) = e^{i (x_0 \hat H_0 + x_1\hat H_1)}~,
\end{equation}
where $x_0=t$ and $x_1=bt$. Employing the above gate in Eq.~\eqref{cost}, from \eqref{leibniz} we get 
\begin{align}
	\partial_t C &= \frac{\partial C}{\partial x_0} + \frac{\partial C}{\partial x_1}b~,
							 &
	\partial_b C &= \frac{\partial C}{\partial x_1}t~.
\end{align}
An estimator of $\frac{\partial C}{\partial x_j}$ for $j=0,1$ can be obtained 
with Algorithm~\ref{alg:sps}, where $\hat V$ is, respectively, either $\hat H_0$ or $\hat H_1$. 
Step 3 in the algorithm corresponds to $\hat U((1-s)t,b)$ 
and Step 5 corresponds to $\hat U(st,b)$, so both operations can be easily implemented directly
in the device. Step 4 corresponds to the gate $U(\pi/4,0)$ when estimating 
$\frac{\partial C}{\partial x_0}$, which is again easy to implement. However, Step 4 
for estimating $\frac{\partial C}{\partial x_1}$ corresponds to the gate $e^{i\pi \hat H_1/4}$ 
that does not belong to the set of gates \eqref{withdrift} and, with our assumptions, 
cannot be implemented by the device. However, we may substitute that gate with an
approximation 
\begin{equation}
	U\left(\epsilon, \frac\pi{4\epsilon}\right) = 
	e^{i \left(\epsilon \hat{H}_0 \pm \frac\pi4 \hat{H_1}\right)}
	= e^{\pm i \frac\pi4 \hat H_1} + \mathcal O(b^{-1}),
 \label{gatedrift}
\end{equation}
where $b=\frac\pi{4\epsilon}$.
The error coming from the drift term can be small $O(\epsilon)$ 
if it is possible to set $b$ to a high value $O(\epsilon^{-1})$.
With the above gate, in Fig.~\ref{fig:asps} we define the approximate \sps.
\begin{figure}[t]
	\begin{algorithm}[H]
		\caption{Approximate \sps }
		\label{alg:asps}
		\begin{algorithmic}[1]
			\State Sample $s$ from the uniform distribution in [0,1];
			\For{$m=\{+,-\}$}
			\State initialize the computer in the state $\ket{\phi}$;
			\State apply  the gate $e^{i (1-s)(\hat H+x \hat V)}$;
			\State apply the gate $e^{i\epsilon [\hat H \pm\pi/(4\epsilon) \hat V]}$
			where the sign 
			\StatexIndent[1.3] depends on $m$;
			\State apply  the gate $e^{i s(\hat H+x \hat V)}$;
			\State measure the observable $\hat A$ and call 
			\StatexIndent[1.3] the result $r_m$.
			\EndFor
			\State An estimate $g_{t,\mb\nu}$ of $\partial C/\partial x_{t,\mb\nu}$ is given by $g_{t,\mb\nu}=r_+-r_-$. 
		\end{algorithmic}
	\end{algorithm}
	\noindent
	\caption{
		Approximate Stochastic Parameter Shift Rule. A compact notation has been used, as 
		this algorithm is identical to the one in 
		Fig.~\ref{fig:sps}, except for the use of the imperfect gates 
		$e^{i\epsilon [\hat H \pm\pi/(4\epsilon) \hat V]}$ in lieu of 
		$e^{\pm \pi/4 \hat V}$.
	}%
	\label{fig:asps}
\end{figure}
The approximate gate introduces a bias in the gradient estimator, but since such bias can be 
made small, convergence can still be expected \cite{spall1992multivariate}.

\begin{figure}[t]
	\centering
	\includegraphics[width=0.9\linewidth]{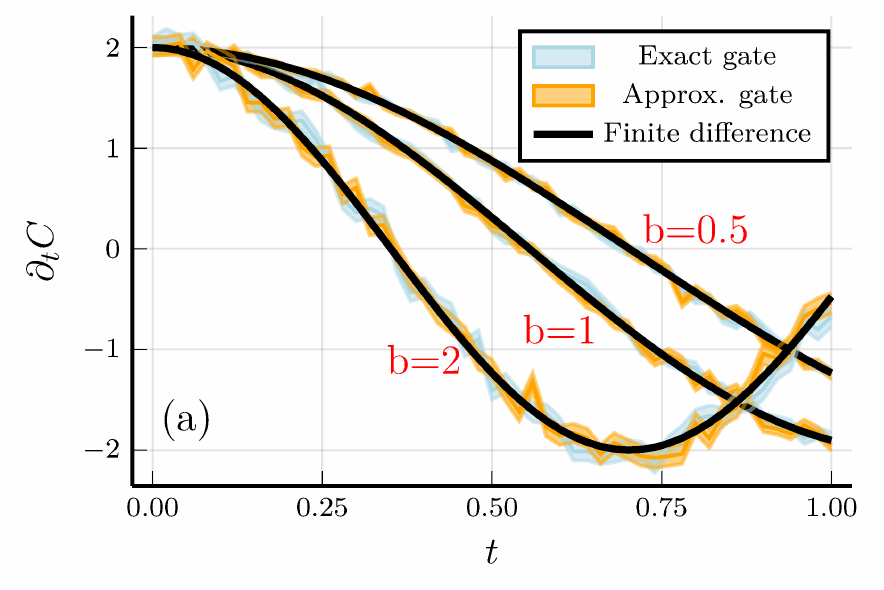}
	\\
	\includegraphics[width=0.9\linewidth]{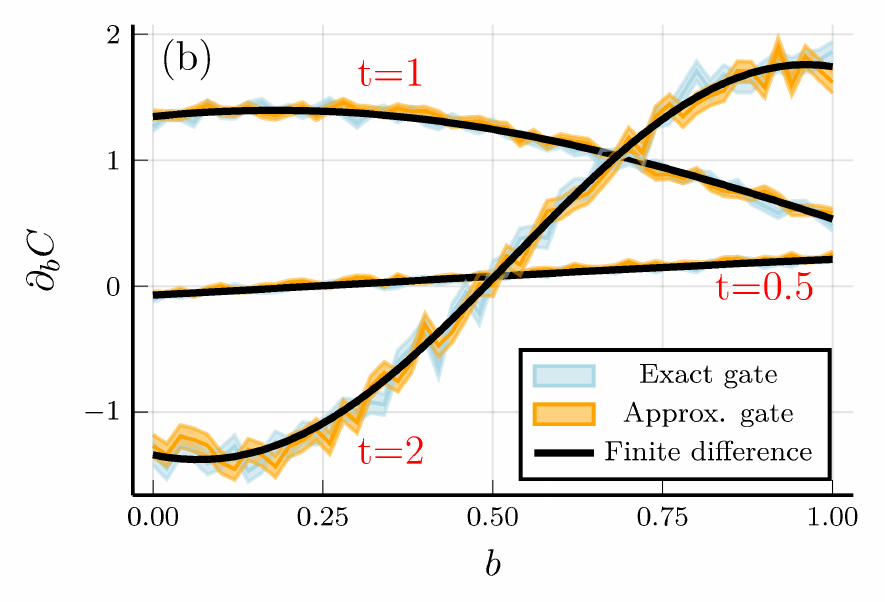}
	\caption{ Gradient of Eq.~\eqref{cost} when using the cross-resonance gate 
		\eqref{xr}. In (a) we study $\partial_t C$ for $c=0$ and fixed values of $b=\{0.5,1,2\}$ 
		with $\hat C = \hat \sigma_y\otimes\hat\openone$. 
		In (b) we study $\partial_b C$ for 
		 $c=\sqrt{2}$ and fixed values of $t=\{0.5,1,2\}$ 
		with $\hat C = \hat \sigma_y\otimes\hat\sigma_y$. We compare the finite difference approximation, 
		with the estimations from Algorithm~\ref{alg:sps} (Exact gate) or \ref{alg:asps} (Approx. gate). 
		Approximated gates are with $\epsilon=10^{-2}$. Data for the stochastic algorithms 
		are obtained from \eqref{stocgrad} with 1000 samples. Coloured regions represent the area $m\pm\sigma$ 
		where $m$ is the estimated mean and $\sigma$ the standard error of the mean. 
	}%
	\label{fig:xr}
\end{figure}

As a relevant example, we study the cross-resonance gate \cite{schuld2019evaluating,crooks2019gradients}
\begin{equation}
	\hat U_{\rm CR}(t,b,c) = \exp\left[it \left(\hat\sigma_x{\otimes}\hat\openone-b\,\hat\sigma_z{\otimes}\hat\sigma_x
		+c\,\hat\openone{\otimes}\hat\sigma_x\right)\right]~,
	\label{xr}
\end{equation}
a natural gate for certain microwave-controlled transmon superconducting qubit
architectures \cite{chow2011simple}. 
The results are shown in Fig.~\ref{fig:xr} for different values of $t$, $c$ and $b$, 
where we show that Algorithms~\ref{alg:sps} and \ref{alg:asps} are basically indistinguishable 
from each other, and very close to the approximated value obtained numerically, without any randomness, 
using a finite difference approximation. 
All numerical results are obtained by 
analytically computing the probabilities \eqref{probms} and then 
simulating the quantum measurement via Monte Carlo sampling. 
The finite difference approximation is obtained as $\partial_x C(x)\approx 
(2\varepsilon)^{-1}[C(x+\varepsilon)-C(x-\varepsilon)]$. Note that, although 
this approximation works fine for numerical approximations using a classical computer, 
it is not useful for calculating gradients on quantum hardware. Indeed, if we 
use a quantum device for estimating $C(x\pm\varepsilon)$, then the estimator of
$\partial C_x$ has a variance $\approx \varepsilon^{-2}$ which is very high when 
$\varepsilon$ is small.

\section{Applications}\label{sec:appl}
\subsection{Quantum control with drift} 
The control of a quantum system is obtained by modulating the interactions 
via time-dependent pulses. Calling $\lambda_j(t)$ the external pulses and $\hat V_j$ 
the associated operators, the evolution is described by the following time-dependent 
Hamiltonian 
\begin{equation}
	\hat H(t) = \hat H_0 + \sum_{j=1}^M \lambda_j(t) \hat V_j~, 
\end{equation}
where $M$ is the number of pulses and $\hat H_0$ is the {\it drift Hamiltonian}
that describes the time-evolution of the system when no pulses are applied. 
Here we consider $M=1$ as the generalization is straightforward, and set $\lambda_1\equiv \lambda$ 
and $\hat V_1\equiv \hat V$. 
By discretizing the control time $T$ into $N_T=T/\Delta T$ steps of width $\Delta t$ we get 
\begin{equation}
	\hat U(T) \approx \prod_{p=1}^{N_T} e^{-i\Delta T(\hat H_0 + \lambda(p \Delta T)\hat V)}~
\end{equation}
with error $\approx N_T\Delta T^2$. Pulse design corresponds to the optimization 
of the parameters $\theta_p := \lambda(p \Delta T)$ to achieve 
a desired target evolution \cite{khaneja2005optimal}, 
for which we can apply the procedure of the section \ref{sec:main_idea}. 
An alternative is to expand the pulse in the Fourier basis 
$\lambda(t) = \sum_m a_m \cos(\omega_k t+\phi_k)$ for some frequencies $\omega_k$ 
and tunable amplitudes $a_m$ and phases $\phi_k$ \cite{caneva2011chopped}. Therefore, we may use 
\eqref{leibniz} together with the procedure of Sec.~\ref{sec:main_idea} 
to estimate the gradient with respect to the parameters $\{\theta_p\} = 
\{a_m,\phi_m\}$. 

\subsection{Parametric circuits with noisy quantum gates}
\label{sec:noisy}

\begin{figure}[t]
	\centering
	\includegraphics[width=0.8\linewidth]{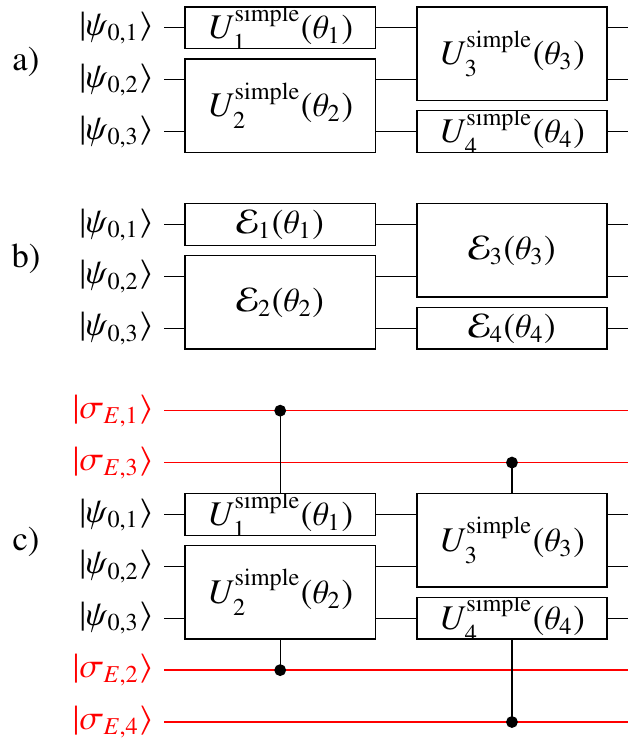}
	\caption{a) An example parametric quantum circuit with parametric gates as in Eq.~\eqref{Usimple}. 
		b) A noisy version of a), where unitary gates are replaced by non-unitary channels. 
		c) A representation \eqref{cpmap} of the noisy gates in b), where each noisy operation 
		is represented as a unitary gate between the qubits and an independent environment (in red). 
	}%
	\label{fig:gates}
\end{figure}

One of the main strengths of our Algorithm~\ref{alg:asps}, and its generalizations in 
Appendix~\ref{a:explicit}, is its ability to 
work, under reasonable assumptions, even when parametric gates are 
not perfectly implemented by the device. This is the case in currently available 
and near-term quantum computers \cite{preskill2018quantum}. 

As a relevant example, 
consider the quantum circuit of Fig.~\ref{fig:gates}, 
built from simple parametric gates as in Eq.~\eqref{Usimple}. 
When the quantum computer can apply the exact gates, then 
the standard parameter shift rule can be employed.
However, quantum devices are always in contact with their surrounding environment, so an exact 
application of the gate is impossible (without full quantum error correction). More precisely, due to the action of the environment 
the gate is not unitary but, under some reasonable approximations, can be described by
a completely positive map \cite{rivas2012open,breuer2002theory}. A completely positive 
map can always be written as a unitary evolution on the register and its environment. 
For simplicity let us consider a perfect gate as in \eqref{Usimple} with fixed $t$. 
Physically, the perfect gate \eqref{Usimple} means that a control  Hamiltonian 
$\hat H^{(R)}_t := -\hat \sigma_{\mb\nu(t)}$ is switched on for a time $\theta_t$, where 
the index (R) reminds us that the Hamiltonian acts on the registers $R$ only. 
In realistic implementations the register is 
coupled with its own environment. If we call $H_t^{(RE)}$ the 
coupling Hamiltonian between register (R) and environment (E), then we may write the 
non-unitary gate (see also Fig.~\ref{fig:gates}c) as 
\begin{equation}
	\mathcal E_t(\theta)[\rho] = \Tr_E\left[e^{-i\tau H_{t,\theta}^{(RE)}}\rho_R\otimes\sigma_E
	e^{i\tau H_{t,\theta}^{(RE)}}\right]~,
	\label{cpmap}
\end{equation}
where $\sigma_E$ is the state of the environment, $\tau$ is the control time, 
$H_{t,\theta}^{(RE)} = H_t^{(RE)}+ \theta H_t^{(R)}$ 
and 
$\theta$ is the relative strength between %the register Hamiltonian
$H_t^{(R)}$ and %and coupling Hamiltonian 
$H_t^{(RE)}$.  In Eq.~\eqref{cpmap} there are three main approximations: 
i) we neglect any initial quantum correlation between register and environment, 
so that the non-unitary evolution can 
be modeled as a completely positive map \cite{rivas2012open}, which in turn implies \eqref{cpmap}; 
ii) we assume that the (unknown) initial state of the environment does not depend on $\theta$ and $\tau$;
iii) we assume that it is possible to tune both $\tau$ and, to some extent, the relative strength~$\theta$. 
Under these three conditions, it is possible to use Algorithm~\ref{alg:asps} and its generalizations 
of Appendix~\ref{a:explicit} to compute the gradient with respect to $\theta$. 
Indeed, without loss of generality, we may consider $H_t^{(R)}$ as a product of Pauli matrices 
acting on the register $R$. When this is not the case we may employ the Leibniz rule 
\eqref{leibniz}. All operations in Algorithm~\ref{alg:asps} are possible, with the 
substitution $\hat H_0=\hat H_t^{(RE)}$ and $ \hat V=\hat H_t^{(R)}$. The rescaled gates 
correspond to reducing the control time $\tau$ by either a factor $(1-s)$ or $s$, 
while the application of the approximate gate~\eqref{gatedrift} can be obtained by making 
$\theta$ large. Note that in a {\it good} quantum computer, the factor $\theta$ should always 
be large, as the coupling between register and environment should be small. 
Therefore, derivatives  with respect to $\theta$ can be obtained using the same operations 
available in the device. 

On the other hand, derivatives with respect to $\tau$ are, in general, not possible. 
We may always expand the coupling Hamiltonian in the Pauli basis via \eqref{expansion}
and use the Leibniz rule \eqref{leibniz},
but in order to obtain the derivative with our Algorithm~\eqref{alg:asps}, 
we have to approximate a highly tuned gate of type 
$e^{i\pi/4 \hat \sigma^{(RE)}}$, which couples the system and environment. 
We believe that for reasonable models of environment, this is not generally possible. 

In summary, when the noisy evolution can be written as in Eq.~\eqref{cpmap},
under the approximations defined above, derivatives with respect to $\theta$ 
can be obtained with the same operations available in the machine, 
while the further parameter $\tau$ should only be used 
to implement the rescaling and not as an optimization parameter.

\subsection{Quantum Natural Gradient} \label{sec:natgrad}
The quantum natural gradient has been proposed in \cite{stokes2019quantum,koczor2019quantum} 
as a way to exploit the geometry of parametric quantum states during optimization,
enabling faster convergence towards local optima. With the quantum natural gradient 
the update rule becomes 
$\mb\theta\to\mb\theta\pm\eta \tilde F^{-1}\mb g$, where $\mb g$ is the 
gradient and $\tilde F$ the metric tensor. The role of the metric 
tensor for noisy parametric quantum evolution has been studied first in \cite{gentini2019noise},
where it was shown that it provides a method to investigate the convergence time 
of standard stochastic gradient descend. 
When using the simple parametric gates 
of \eqref{Usimple}, the elements of this tensor can be measured efficiently for pure states
\cite{stokes2019quantum,yuan2019theory}. 
Moreover, recently the quantum natural gradient has been extended to 
arbitrary noisy quantum states \cite{koczor2019quantum}. In particular, for 
slightly mixed states it is 
\begin{align}
	\tilde F_{p,p'} &\approx \kappa F_{p,p'}, &F_{p,p'}:=\Tr\left[\frac{\partial \hat \rho}{\partial \theta_p}
	\frac{\partial \hat \rho}{\partial \theta_{p'}}\right]~,
	\label{simon}
\end{align}
where $\kappa=1$ for pure states and $\hat\rho$ is the state after the parametric unitaries that, 
for either noiseless or noisy gates, we can write as 
$\hat\rho(\mb\theta) = \mathcal E_T(\mb\theta)\circ \dots\circ \mathcal E_1(\mb\theta)[\hat\rho_0]$ with 
$\hat\rho_0=\ket{\psi_0}\!\bra{\psi_0}$. 
We focus on $F_{p,p'}$ as the parameter $\kappa$ can be absorbed into the learning rate. 
The approximation in \eqref{simon} is valid when the state 
has a high purity \cite{koczor2019quantum}, as it is expected in {\it good} NISQ computers. 
We may measure the matrix in Eq.~\eqref{simon} using a 
combination of the \sps~and the SWAP test. The latter is based 
on the simple observation that, for any $\hat X$ and $\hat Y$, it is 
$\Tr[\hat X\hat Y]=\Tr[\hat S(\hat X\otimes\hat Y)]$,
where $\hat S$ is the swap operator \cite{yuan2019theory}.
Using the SWAP test and Eq.~\eqref{leibniz} we get %may write 
\begin{equation}
	F_{p,p'} = \sum_{t,t',\mb\nu,\mb\nu'} 
	\Tr\left[ \hat S \left(
		\frac{\partial \hat \rho}{\partial x_{t,\mb\nu}}\otimes
	\frac{\partial \hat \rho}{\partial x_{t',\mb\nu'}}\right)\right]
	\frac{\partial x_{t,\mb\nu}}{\partial \theta_p}
	\frac{\partial x_{t',\mb\nu'}}{\partial \theta_p'}~.
\end{equation}
Then, thanks to our analysis from section~\ref{sec:main_idea}, we may write 
\begin{align}
	\label{metric}
	&F_{(t,\mb\nu),(t',\mb\nu')} := \Tr\left[ \hat S\left(
		\frac{\partial \hat \rho}{\partial x_{t,\mb\nu}}\otimes
	\frac{\partial \hat \rho}{\partial x_{t',\mb\nu'}}\right)\right]
														= \\&~~=\nonumber 
														\!\!\!\!\!\!\!
														\sum_{\alpha=\pm,\alpha'=\pm} 
														\!\!\!\!\!\!\!
														\alpha\alpha'
\int_0^1 ds\int_0^1 ds' \Tr[\hat S \left(\hat	\rho_{t,\mb\nu,s,\alpha} \otimes
														\hat	\rho_{t',\mb\nu',s',\alpha'}\right) ],
\end{align}
where $\hat \rho_{t,\mb\nu,s,\pm} $ is the state in which the gate $\hat U_t$ has been substituted 
by the gate $U_{\pm}(x_{t,\mb\nu},s)$ from Eq.~\eqref{Upm}, or its noisy implementation as 
in Sec.~\ref{sec:noisy}. 
Therefore, an estimator of the matrix elements of the Fisher information matrix can be obtained 
by sampling two real numbers $s$ and $s'$ from the uniform distribution, and then measuring 
the overlaps of all quantum states 
$	\rho_{t,\mb\nu,s,\alpha} $ and  $\rho_{t',\mb\nu',s',\alpha'} $ via the swap test.
Note that for noiseless gates the overlaps in \eqref{metric} can be simplified in some 
cases. For instance, when $t'=t$ all the gates in the product \eqref{Uprod} with larger~$t$ 
disappears from the overlap. 
It was found in \cite{stokes2019quantum} that a good approximation to the natural gradient 
can be obtained by using only the diagonal elements of $F$. Motivated by this, we study 
what happens when we fix $t$ and~$\mb\nu$ and call $x\equiv x_{t,\mb\nu}$ 
as in Sec.~\ref{sec:main_idea}. 
With the notation of Eqs.~\eqref{prep} and \eqref{cost_simple},  
using \eqref{id}
%and setting $\hat\rho = e^{\mathcal Z}[\hat\rho_0]$ 
we may write
\begin{align}
	F_{(t,\mb\nu),(t,\mb\nu)} &= \nonumber
	\Tr[\frac{\partial \hat \rho}{\partial x}\frac{\partial \hat \rho}{\partial x}] 
	=
	\Tr(\frac{\partial e^{\mathcal Z}}{\partial x}[\hat\rho_0]
	\frac{\partial e^{\mathcal Z}}{\partial x}[\hat\rho_0]) = 
%													\\&= \int_0^s ds\int_0^1ds' 
%	\Tr\left[e^{s \mathcal Z} \frac{\partial \mathcal Z}{\partial x} 
%	e^{(1-s) \mathcal Z}[\hat\rho_0]
%	e^{s' \mathcal Z} \frac{\partial \mathcal Z}{\partial x} 
%	e^{(1-s') \mathcal Z}[\hat\rho_0]
%	\right]~.
													\\&= \int_0^1ds\int_0^1 ds'\Tr(
													i[\hat V(s),\hat\rho_0]
													i[\hat V(s'),\hat\rho_0]
													)
													\nonumber = 
													\\&= 2(F_2-|F_1|^2)~,
\end{align}
where $\hat\rho_0=\ket\phi\!\bra\phi$, 
$\hat V(s)=e^{is(\hat H+x\hat V)}\hat V
e^{i(1-s)(\hat H+x\hat V)}$  and we have defined 
\begin{align}
	F_2 &= \int_0^1ds\int_0^1ds' \bra\phi \hat V(s)\hat V(s')\ket\phi~,
	\\
	F_1 &= \int_0^1ds \bra\phi \hat V(s)\ket\phi~.
\end{align}
Since $\hat V$ is a product of Pauli matrices  $\hat V(s)$ is a 
unitary operator, so both $F_2$ and $F_1$ can be measured by first 
sampling $s$ and $s'$ from 
the uniform distribution, and then measuring the expectation value 
using the Hadamard test \cite{mitarai2019methodology}.

\section{Conclusions}\label{sec:conc}

We have studied the optimization of a {\it cost function} defined by taking a quantum measurement 
on a parametric quantum state, obtained by applying on a fixed reference state a controlled 
evolution with tunable classical parameters. 
We have found explicit analytical formulae for the derivatives of the cost function with 
respect to those classical parameters. Our formulae can be applied to any 
multi-qubit evolution and generalize the so-called parameter shift rule
\cite{mitarai2018quantum,schuld2019evaluating} to the general case, without any
restriction on the spectrum of the operator, and without the use of ancillary qubits 
or Hamiltonian simulation techniques \cite{childs2012hamiltonian}. 

Based on those exact formulae, we have devised both exact and approximate 
algorithms for estimating the derivatives of the cost via carefully designed 
quantum circuits. The exact algorithm works when exact applications of the gates
are possible, whereas the approximate algorithm is  designed to tackle spurious
interactions in the system that cannot be completely removed. As such, 
our algorithm can also be applied, though with some approximations, when
the gates implemented by the quantum device are noisy, as it is the 
case in near-term quantum devices \cite{preskill2018quantum}.

The main application of our study is to optimize parametric quantum evolution 
for quantum optimization \cite{peruzzo2014variational} and machine-learning 
problems \cite{schuld2018supervised}. 

\begin{acknowledgements}
	The authors acknowledge Xanadu Inc.~for hosting the QHACK'19 event 
	(\url{https://qhack.ai/}) where this study was initiated. 
	L.B. acknowledges support by the program ``Rita Levi Montalcini'' for young researchers. X, formerly known as Google[x], is part of the Alphabet family of companies, which includes Google, Verily, Waymo, and others (\url{www.x.company}).
\end{acknowledgements}

\appendix

\section{Hadamard test} \label{a:hadamard}
We discuss how the  Hadamard test
\cite{mcclean2018barren,harrow2019low,yuan2019theory,mitarai2019methodology} can be formally written 
for general gates. While it is possible to formally write the gradient in the general case, 
we show that its efficient estimation via the Hadamard test may be limited to gates acting on few qubits. 
Without loss of generality, we consider the cost \eqref{cost_simple} and focus on $\hat U(x)=e^{i(\hat H+x \hat V)}$. 
Thanks to the identity \eqref{id} we can write
\begin{equation}
	\frac{\partial \hat U(x)}{\partial x} = i\hat Y(x)\hat U(x)~,
\end{equation}
where
\begin{equation}
	\hat Y(x) =	\int_0^1  e^{is(\hat H+x \hat V)}\hat V 
	 e^{-is(\hat H+x \hat V)}\,ds~,
	 \label{Yinte}
\end{equation}
and from Eq.~\eqref{cost_simple} 
\begin{equation}
	\frac{\partial C(x)}{\partial x} = i \bra\phi \hat U(x)^\dagger [\hat A,\hat Y(x)] \hat U(x) \ket\phi.
	\label{cost_hada}
\end{equation}
Since $\hat Y(x)$ is a Hermitian operator, it can be expanded in the Pauli basis as 
\begin{equation}
	\hat Y(x) = \sum_{\mb\mu} y_{\mb\mu}(x) \sigma_{\mb\mu}.
	\label{ysum}
\end{equation}
Inserting the above expansion in \eqref{cost_hada}, we see that the Hadamard test can be extended to 
the general case, provided that the expansion coefficients $y_{\mb\mu}(x)$ are efficiently computable and 
the number of non-zero terms in the sum \eqref{ysum} is sufficiently small. 
This is the case when the operators $\hat H$ and $\hat V$ act non-trivially on a few qubits but, in general,
the formal solution of Eq.~\eqref{Yinte}, or other methods with comparable complexity, is too difficult 
for generic many-body operators $\hat H$ and $\hat V$. Moreover, for generic many-body operators,
the number of non-zero terms in \eqref{ysum} is 
expected to grow exponentially with the number of qubits.

\section{Explicit algorithms} \label{a:explicit}

In this appendix we discuss more explicitly all the steps to 
define unbiased estimators of $\partial C/\partial \theta_p$ that 
can be measured with the~\sps.
The full version of Algorithms~\ref{alg:sps} and \ref{alg:asps} 
is the following: 

\begin{algorithm}[H]
	\caption{Stochastic Parameter Shift Rule, Eq.~\eqref{stocgrad} }
  \label{alg:stocgrad}
   \begin{algorithmic}[1]
		 \State Sample $s$ from the uniform distribution in $[0,1]$;
		 \State set $g_p=0$;
		 \For{$t=1,\dots,T$}
		 \ForAll{$\mb\nu$ such that $\partial_{\theta_p} x_{t,\mb\nu}(\mb\theta)\neq 0$}
		 \For{$m=\{+,-\}$}
		 \State initialize the computer in the state $\ket{\psi_0}$;
		 \State sequentially apply the gates $\hat U_{t'}$ for 
		 \StatexIndent[3.3] $t'=0\dots,t-1$ to prepare the state $\ket{\phi}$ 
		 \StatexIndent[3.3] in Eq.~\eqref{prep};
		 \State apply  the gate $\hat U_t^{1-s}\equiv e^{i (1-s)\hat X_t(\mb\theta)}$, by 
		 \StatexIndent[3.3] rescaling all  parameters;
		 \If{gates $e^{\pm i\frac\pi4 \hat\sigma_{t,\mb\nu}}$ are available}{}
		 \State apply the gate $e^{m i \frac\pi4  \hat \sigma_{t,\mb\nu}}$;
%		 where the sign \StatexIndent[4.3] depends on $m$;
		 \Else
		 \State apply the best approximation of 
		 \StatexIndent[4.3] $e^{m i \frac\pi4 \hat \sigma_{t,\mb\nu}}$, for instance using \eqref{gatedrift}; 
		 \EndIf
		 \State apply  the gate $\hat U_t^{s}\equiv e^{i s\hat X_t(\mb\theta)}$;
		 \State sequentially apply the gates $\hat U_{t'}$ for 
		 \StatexIndent[3.3] $t'=t+1,\dots,T$;
		 \State measure the observable $\hat C$ and call 
		 \StatexIndent[3.3] the result $r_{m,t,\mb\nu}$;
		 \EndFor
		 \State set  $g_{t,\mb\nu}=r_{+,t,\mb\nu}-r_{-,t,\mb\nu}$;
		 \State update $g_p\to g_p + g_{t,\mb\nu} \partial_{\theta_p} x_{t,\mb\nu}(\mb\theta)$
		 \EndFor
		 \EndFor
		 \State the sample $g_p$ is such that $\partial C /\partial \theta_p = \mathbb{E}[g_p]$.
   \end{algorithmic}
\end{algorithm}
By repeating the analysis of Sec.~\ref{sec:stoc_opt} we find that 
$\partial C /\partial \theta_p = \mathbb{E}[g_p]$, so by repeating Algorithm~\ref{alg:stocgrad} 
many times we may estimate the derivative 
$\partial C/\partial\theta_p$ with the desired precision. A simple 
counting argument shows that the number of operations to obtain a single 
outcome is $\mathcal O(2TN_p)$ where $N_p$ is the number of non-zero 
$\partial_{\theta_p} x_{t,\mb\nu}(\mb\theta)$. 
Note that Step 1: in Algorithm \ref{alg:stocgrad} can be moved 
to any other point point before Step 8. By linearity, the average is always the
same, although each iteration might have a different value of $s$. 
We can reduce the number of operations to get a single estimate with the
following algorithm:

\begin{algorithm}[H]
	\caption{Doubly Stochastic Parameter Shift Rule, Eq.~\eqref{dstocgrad} }
  \label{alg:dstocgrad}
   \begin{algorithmic}[1]
		 \State Sample $s$ from the uniform distribution in $[0,1]$;
		 \State calculate the probability distribution $q_p(t,\mb\nu)$ defined in Eq.~\eqref{probtnu} 
		 and set $n_{p,t,\mb\nu}$ as described in Sec.~\ref{sec:stoc_opt}; 
		 \State sample $(t,\mb\nu)$ from $q_p$;
		 \For{$m=\{+,-\}$}
		 \State initialize the computer in the state $\ket{\psi_0}$;
		 \State sequentially apply the gates $\hat U_{t'}$ for 
		 \StatexIndent[1.3] $t'=0\dots,t-1$ to prepare
		 the state $\ket{\phi}$ 
		 \StatexIndent[1.3] in Eq.~\eqref{prep};
		 \State apply  the gate $\hat U_t^{1-s}\equiv e^{i (1-s)\hat X_t(\mb\theta)}$, by 
		 \StatexIndent[1.3] rescaling all parameters;
		 \If{gates $e^{\pm i\frac\pi4 \hat\sigma_{t,\mb\nu}}$ are available}{}
		 \State apply the gate $e^{m i \frac\pi4  \hat \sigma_{t,\mb\nu}}$;
		 \Else
		 \State apply the best approximation of $e^{m i \frac\pi4 \hat \sigma_{t,\mb\nu}}$, 
		 \StatexIndent[4.3] for instance using \eqref{gatedrift}; 
		 \EndIf
		 \State apply  the gate $\hat U_t^{s}\equiv e^{i s\hat X_t(\mb\theta)}$;
		 \State sequentially apply the gates $\hat U_{t'}$ for 
		 \StatexIndent[1.3] $t'=t+1,\dots,T$;
		 \State measure the observable $\hat C$ and call the result 
		 \StatexIndent[1.3] $r_{m,t,\mb\nu}$;
		 \EndFor
		 \State the sample $g_p = (r_{+,t,\mb\nu}-r_{-,t,\mb\nu})n_{p,t,\mb\nu}$ is such that 
		 $\partial C /\partial \theta_p = \mathbb{E}[g_p]$.
   \end{algorithmic}
\end{algorithm}
In Algorithm~\eqref{alg:dstocgrad} the quantum computer is still reset twice to have a single 
estimate. Below we define an algorithm where the computer is initialized only once 

\begin{algorithm}[H]
	\caption{Single-measurement sample of $\partial C/\partial\theta_p$}
  \label{alg:ddstocgrad}
   \begin{algorithmic}[1]
		 \State Sample $s$ from the uniform distribution in $[0,1]$;
		 \State calculate the probability distribution $q_p(t,\mb\nu)$ defined in Eq.~\eqref{probtnu} 
		 and set $n_{p,t,\mb\nu}$ as described in Sec.~\ref{sec:stoc_opt}; 
		 \State sample $(t,\mb\nu)$ from $q_p$;
		 \State initialize the computer in the state $\ket{\psi_0}$;
		 \State sequentially apply the gates $\hat U_{t'}$ for $t'=0\dots,t-1$ 
		  to prepare the state $\ket{\phi}$ in Eq.~\eqref{prep};
		 \State apply  the gate $\hat U_t^{1-s}\equiv e^{i (1-s)\hat X_t(\mb\theta)}$, by rescaling all 
		 parameters;
		 \State sample $m\in\{+1,-1\}$ by tossing a fair coin;
		 \If{gates $e^{\pm i\frac\pi4 \hat\sigma_{t,\mb\nu}}$ are available}{}
		 \State apply the gate $e^{m i \frac\pi4  \hat \sigma_{t,\mb\nu}}$;
		 \Else
		 \State apply the best approximation of $e^{m i \frac\pi4 \hat \sigma_{t,\mb\nu}}$, 
		 \StatexIndent[1.3] e.g. using \eqref{gatedrift}; 
		 \EndIf
		 \State apply  the gate $\hat U_t^{s}\equiv e^{i s\hat X_t(\mb\theta)}$;
		 \State sequentially apply the gates $\hat U_{t'}$ for $t'=t+1,\dots,T$;
		 \State measure the observable $\hat C$ and call the result $r$;
		 \State the sample $g_p = 2 m r n_{p,t,\mb\nu} $  is such that $\partial C /\partial \theta_p = \mathbb{E}[g_p]$.
   \end{algorithmic}
\end{algorithm}

The above algorithm corresponds to rewriting Eq.~\eqref{cost_deriv} as 
	\begin{equation}
		\partial_x C(x) = \sum_{m=\pm } m p_m \int_{0}^1 2 C_m(x,s)\,ds~.
	\label{cost_deriv2}
\end{equation}
with probabilities
$	p_\pm = \frac12$.
Putting explicitly the dependence on $t$ and $\mb\nu$ we get from \eqref{dstocgrad} 
and from the notation \eqref{xdef}, \eqref{prep}
\begin{align}
	k\frac{\partial C(\mb\theta)}{\partial \theta_p}  = \sum_{t,\mb\nu,m} &m n_{p,t,\mb\nu} p_m q_p(t,\mb\nu)
	\times\\&\times	\int_0^1 2 \bra{\psi_{t,\mb\nu,s,m}}\hat C
	\ket{\psi_{t,\mb\nu,s,m}}
	ds, \nonumber
\end{align}
where $m=\pm$, $t=1,\dots,T$, and 
\begin{equation}
	\ket{\psi_{t,\mb\nu,s,m}} := \prod_{t'=t+1}^T \hat U_{t'} 
	\hat U_t^{1-s} e^{mi\pi/4 \hat \sigma_{\mb\nu}} \hat U_t^s
	\prod_{t'=1}^{t-1} \hat U_{t'} \ket{\psi_0}~.
\end{equation}
It is worth noting that, depending on structure of the observable $\hat C$, the number of 
measurements may be reduced by optimally distributing the number of shots 
\cite{romero2018strategies,van2020measurement} or employing variance reduction 
techniques \cite{gentini2019noise}.

\section{Variance of gradient estimators}\label{s:variance}
By comparing the standard Parameter Shift Rule (Fig.~\ref{fig:ps}) and \sps~(Fig.~\ref{fig:sps}) 
we see that the latter has an extra source of 
randomness due to the sampling over the classical parameter $s$. 
The stochastic outcomes of these two algorithms have the same mean, namely the gradient 
of the cost function, so in the limit of infinitely-many repetitions 
of the experiment these algorithms provide the same result. 
However, the variance of the estimators obtained with the two algorithms 
might be different. Assuming independent identically distributed samples, 
the variance quantifies 
the expected error when a {\it finite} number of measurements 
is performed, so it is important to study whether the extra stochasticity of 
the \sps~increases the variance of the gradient estimators.

\begin{figure}[tp!]
	\centering
	\includegraphics[width=0.9\linewidth]{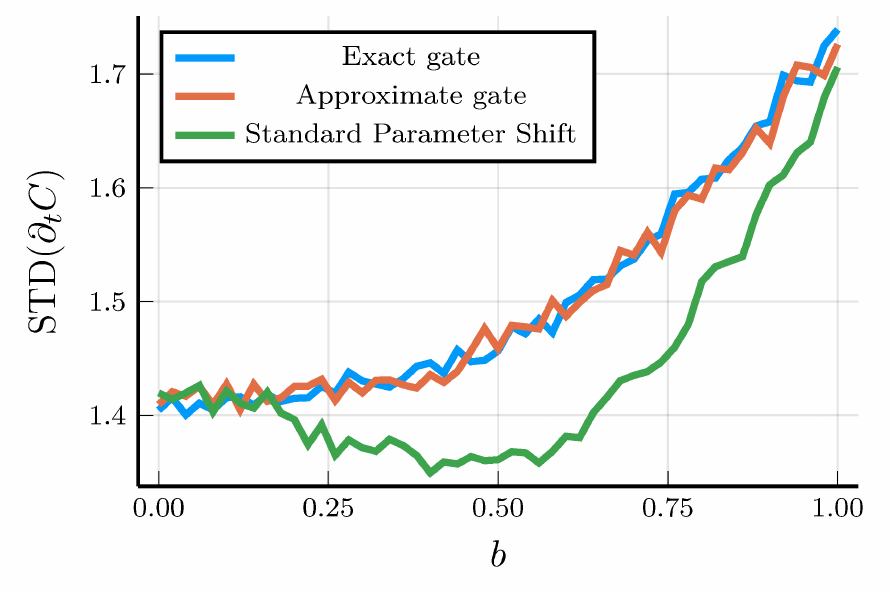}
	\caption{
		Empirical standard deviation of the gradient estimator of $\partial_t C(t,b)$, 
		with the same notation of Fig.~\ref{fig:xr}(b), for $c=0$ and different values of $b$. 
		The Standard Parameter Shift Rule corresponds to Algorithm~\ref{alg:ps}.
		For each point, the STD is estimated using $10^4$ samples. 
	}%
	\label{fig:stdps}
\end{figure}

\begin{figure}[tph!]
	\centering
	\includegraphics[width=0.9\linewidth]{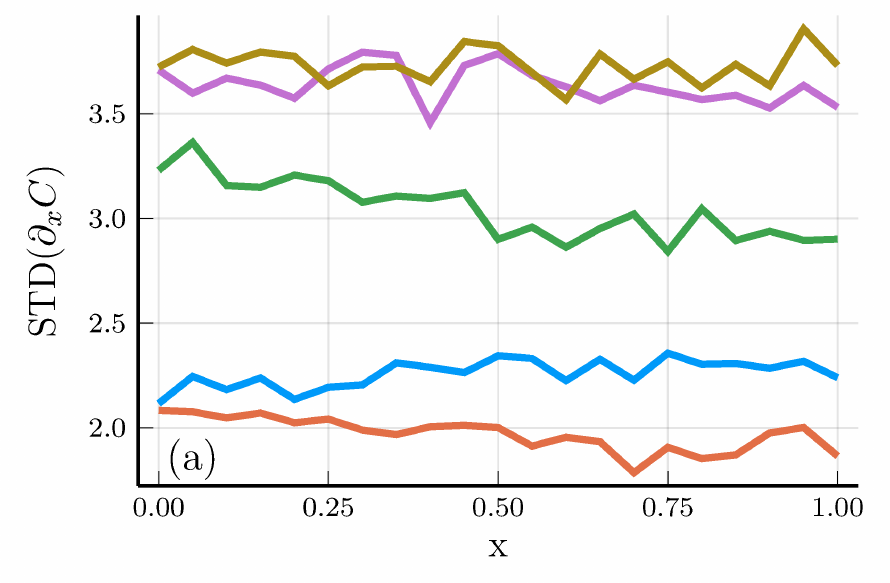}
	\\
	\includegraphics[width=0.9\linewidth]{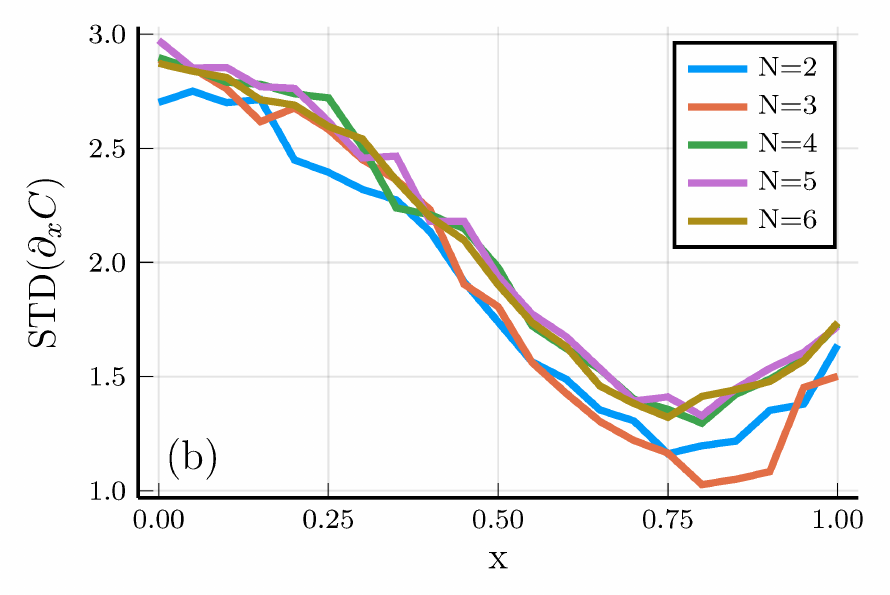}
	\caption{(a) 
		Empirical standard deviation (STD) of the estimator from Algorithm~\ref{alg:sps},
		using $\hat H=\hat H_a$ and 
		$\hat V=\hat V_a$ from
		Eqs.~\eqref{modelA}, for different numbers of qubits $N$.
		(b) 
		STD of the gradient estimator obtained via the (standard) Parameter Shift Rule, 
		for a related problem with $\hat H=0$, shown in Eq.~\eqref{modelB}.
		In both (a) and (b) the plots are shown for different 
		values of the parameter $x$, as in Eq.~\eqref{cost_simple},
		while the STD is estimated via 1000 samples. 
	}%
	\label{fig:ising}
\end{figure}

\begin{figure}[tph!]
	\centering
	\includegraphics[width=0.9\linewidth]{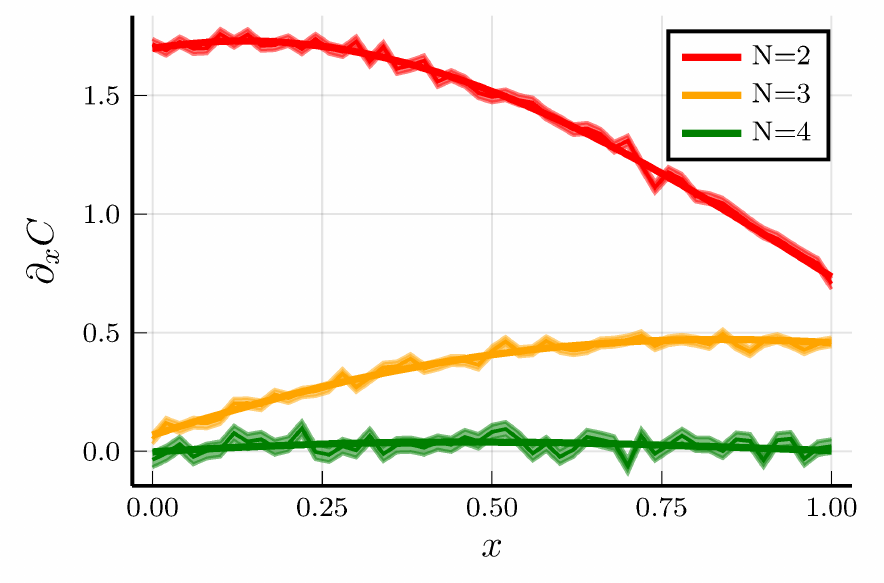}
	\caption{ Finite difference approximation (solid lines), versus estimated gradient 
		via Algorithm~\ref{alg:sps}, with error bars as in Fig.~\ref{fig:xr}. We focus on $C(x)$ with 
		the definitions \eqref{cost_simple} and model \eqref{modelA}, for different numbers of 
		qubits $N$. 
	}%
	\label{fig:manybody}
\end{figure}

We first study the gradient of $C(t,b)$ obtained with $\hat C=\hat\sigma_y\otimes\hat\sigma_y$ 
and the cross-resonance gate \eqref{xr}, as in Fig.~\ref{fig:xr}, but with $c=0$. 
When $c=0$ the operator in the exponential has two possible eigenvalues $u=\pm\sqrt{1+b^2}$, so for 
computing the derivative $\partial_t C$ we can also apply the standard parameter shift rule, 
% simpler rule of Fig.~\eqref{fig:ps}, 
and compare the variance of the resulting estimator with that obtained from the 
\sps. 
Note that, unlike our Algorithm~\ref{alg:sps}, 
the simpler parameter shift rule cannot be applied to estimate $\partial_b C$. 

In Fig.~\ref{fig:stdps} we compare the standard deviation of 
Algorithms~\ref{alg:ps}, \ref{alg:sps}, \ref{alg:asps}. We note that,
although Algorithms~\ref{alg:sps}, \ref{alg:asps} have extra sampling steps,
the resulting variance is comparable with that of Algorithm~\ref{alg:ps}.

We now study how the standard deviation might scale as a function of the number of qubits. 
%In Fig~.\ref{fig:ising} we study the standard deviation of the gradient
%estimator $r_+-r_-$ obtained with a particular model. 
In Fig.~\ref{fig:ising}(a) 
we focus on the \sps, with the following choice of states and operators  
in \eqref{cost_simple}
\begin{subequations}
\begin{align}
	\hat H_a &= \sum_{j=1}^N \left[\hat \sigma_x^{(j)}\hat\sigma_x^{(j+1)}+
	\frac{\hat \sigma_x^{(j)}}3 + \frac{\hat \sigma_z^{(j)}}2\right]~, & 
	\hat V_a &= \sigma_z^{(1)}~,
	\\
	\hat A_a &= \sum_{j=1}^N \hat \sigma_z^{(j)}~, &
	\ket{\phi_a}&=\ket{0}^{\otimes N}, 
\end{align}
\label{modelA}%
\end{subequations}
where $\hat\sigma_x^{(j)}$ means that the operator $\hat \sigma_x$ is applied to the 
$j$the qubit and 
$\hat\sigma_x^{(N+1)} \equiv \hat\sigma_x^{(1)}$. In Eqs.~\eqref{modelA} we have 
chosen for $\hat H_a$ a many-body Hamiltonian with complex entangling dynamics 
\cite{kim2013ballistic}. The empirical mean is shown in Fig.~\ref{fig:manybody} 
for $N=2,3,4$. Larger values of $N$ are not shown, as they are similar to the case $N=4$.
Since $\hat H_a\neq 0$ we cannot apply 
the standard parameter shift rule of Fig.~\ref{fig:ps}. 
Therefore, to compare the algorithms~\ref{alg:ps} and \ref{alg:sps} we 
need to introduce another model with $\hat{H}=0$, namely where all gates depend 
on the parameters as in Eq.~\ref{Usimple}. 
We build such a model using the same operators introduced in Eqs.~\eqref{modelA} and 
define 
\begin{subequations}
\begin{align}
	\hat H_b &= 0~, &
	\hat V_b &= \sigma_z^{(1)}~,
	\\
	\hat A_a &= \sum_{j=1}^N e^{i\hat H_a/2} \hat \sigma_z^{(j)} e^{i\hat H_a/2} ~, &
	\ket{\phi_b}&= e^{i\hat H_a/2} \ket{0}^{\otimes N}.
\end{align}
\label{modelB}%
\end{subequations}
By comparing Fig.~\ref{fig:ising}(a) with Fig.~\ref{fig:ising}(b), we note that the 
standard deviations of both estimators have the same order of magnitude, that does 
not seem to increase too much with the number of qubits $N$, at least for our choice of Hamiltonians.
In Fig.~\ref{fig:ising}(a) we observe a slight non-monotonic increase, while 
in Fig.~\ref{fig:ising}(b) the results are basically independent on $N$. 
We believe that this difference is mostly due to the particular choice of the 
models, Eqs.~\eqref{modelA} and \eqref{modelB}, that although related are 
not identical. 
Therefore, in our numerical studies the stochastic parameter shift rule is basically 
as efficient as the standard parameter shift rule,  but it is more general.

\end{document}